\documentclass[conference]{IEEEtran}
\IEEEoverridecommandlockouts
\usepackage{cite}
\usepackage{amsmath,amssymb,amsfonts}
\usepackage{algorithmic}
\usepackage{graphicx}
\usepackage{xcolor}
\usepackage{multirow}
\newcommand{\tabitem}{~~\llap{\textbullet}~~}
\usepackage{booktabs,threeparttable}
\usepackage{textcomp}
\usepackage{xcolor}
\def\BibTeX{{\rm B\kern-.05em{\sc i\kern-.025em b}\kern-.08em
    T\kern-.1667em\lower.7ex\hbox{E}\kern-.125emX}}
\begin{document}

\title{Large-vocabulary Audio-visual Speech Recognition in Noisy
Environments\\

\thanks{This project has received funding from the German Research Foundation DFG under grant number KO3434/4-2.}
}

\IEEEoverridecommandlockouts
\IEEEpubid{
    \hspace{1mm}
    \makebox[\columnwidth]{978-1-6654-3288-7/21/\$31.00~\copyright 2021 European Union \hfill}
    \hspace{\columnsep}
    \makebox[\columnwidth]{ }
}

\author{\IEEEauthorblockN{Wentao Yu, Steffen Zeiler, Dorothea Kolossa}
\IEEEauthorblockA{\textit{Institute of Communication Acoustics} \textit{Ruhr University Bochum} Germany \\
\{wentao.yu, steffen.zeiler, dorothea.kolossa\}@rub.de}}

\maketitle
\IEEEpubidadjcol
\begin{abstract}
Audio-visual speech recognition (AVSR) can effectively and significantly improve the recognition rates of small-vocabulary systems, compared to their audio-only counterparts. For large-vocabulary systems, however, there are still many difficulties, such as unsatisfactory video recognition accuracies, that make it hard to improve over audio-only baselines. In this paper, we specifically consider such scenarios, focusing on the large-vocabulary task of the LRS2 database, where audio-only performance is far superior to video-only accuracies, making this an interesting and challenging setup for multi-modal integration. 

To address the inherent difficulties, we propose a new fusion strategy: a recurrent integration network is trained to fuse the state posteriors of multiple single-modality models, guided by a set of model-based and signal-based stream reliability measures. During decoding, this network is used for stream integration within a hybrid recognizer, where it can thus cope with the time-variant reliability and information content of its multiple feature inputs. 

We compare the results with end-to-end AVSR systems as well as with competitive hybrid baseline models, finding that the new fusion strategy shows superior results, on average even outperforming oracle dynamic stream weighting, which has so far marked the---realistically unachievable---upper bound for standard stream weighting. Even though the pure lipreading performance is low, audio-visual integration is helpful under all---clean, noisy, and reverberant---conditions. On average, the new system achieves a relative word error rate reduction of 42.18\% compared to the audio-only model, pointing at a high effectiveness of the proposed integration approach.
\vspace{-0.1mm}
\end{abstract}


\section{Introduction}
\label{introduction}
Audio-visual speech recognition (AVSR) is motivated by the natural ability of humans to integrate cross-modal information. When people are listening to speech in a noisy environment, they often unconsciously focus on the speaker\textquotesingle s lips, which is of great benefit to human listening and comprehension~\cite{crosse2016eye}. Even in clean speech, seeing the lips of the speaker influences perception, as demonstrated by the McGurk effect~\cite{mcgurk1976hearing}. It has been shown in many studies~\cite{Wand2017,meutzner2017improving, gurban2008dynamic} that machine AVSR systems can also successfully improve performance on small-vocabulary tasks, when compared to their audio-only speech recognition (ASR) counterparts with otherwise equivalent setups. However, large-vocabulary tasks are still difficult for lipreading, because many phoneme pairs correspond to identical visemes, which makes certain words virtually indistinguishable to a vision-only system, as for example "do" and "to". 

This problem also leads to an inherent difficulty of AVSR on large-vocabulary tasks~\cite{thangthai2018building, sterpu2020}, which is acerbated by the fact that many multi-stream fusion approaches perform badly, when the performance of the streams varies widely. In this work, we address this shortcoming by introducing a new stream fusion strategy that is impervious to such disparate single-stream recognition rates and can still benefit from low-quality streams in improving the results of highly reliable, clean audio data. To evaluate it in a realistic manner, we use a large-vocabulary dataset---the Lip Reading Sentences (LRS2) corpus~\cite{Afouras2018}---for all experiments, which we further augment by adding realistic noise and reverberation. 

An effective fusion strategy for AVSR is decision fusion, which combines the decisions of multiple classifiers into a common decision. Decision fusion comes in different forms, such as dynamic stream-weighting~\cite{stewart2013robust}, or state-based decision fusion (SBDF), e.g.~in \cite{Abdelaziz2015, potamianos2003recent, luettin2001asynchronous, nefian2002dynamic}. An alternative fusion approach is the idea of fusing \emph{representations} rather than decisions, e.g. via multi-modal attentions~\cite{Zhou2019}. Another example in this direction is that of gating, e.g.~in~\cite{Yu2020Overlapped} or in~\cite{arevalo2020gated}, where a newly designed \emph{Gated Multimodal Unit} is used for dynamically fusing feature streams within each cell of a network. 

In this work, we argue that the ideas of representation fusion and decision fusion can be unified in a different fashion, namely, by using the posterior probabilities $p(\textbf{s} |\textbf{o}_t^{\mathrm{i}})$ of $i = 1 \ldots M$ single-modality hybrid models as our representation of the uni-modal streams.

This viewpoint opens up a range of new possibilities, centered around these single-modality representations. On the one hand, we can base the multi-modal model on pre-trained hybrid ASR models. On the other hand, we can learn recurrent and dynamic fusion networks, which can benefit from the reliability information that is inherent in the posterior probabilities, such as instantaneous entropy and dispersion~\cite{gurban2008dynamic}, as well as from temporal context.

Overall, in the following, we compare our new approach with the performance of 4 baseline and oracle fusion strategies, which are detailed in Section~\ref{related}. The proposed fusion strategy is introduced in Section~\ref{systemoverview}. Section~\ref{reliabilitymeasures} describes the set of reliability measures that are employed in all of the dynamic fusion approaches. The experiments are presented in Section~\ref{setup}, while Section~\ref{results} introduces and analyzes the results. Finally, in Section~\ref{conclusion}, we discuss the lessons learned and give an outlook on future work.
\section{Related work}
\label{related}
There are many different fusion strategies  in AVSR research. In this section, we give a brief introduction to the fusion strategies that are considered as baseline models in this work. In these baselines as well as in our own model, M = 3 single-modality models are combined, one acoustic and two visual, where $\textbf{\textrm{o}}_t^\mathrm{A}$ are our audio features, and $\textbf{\textrm{o}}_t^\mathrm{VS}$ and $\textbf{\textrm{o}}_t^\mathrm{VA}$ are shape-based  and appearance-based video features; see Section~\ref{features} for details. 

\subsection{Early integration}
Early integration simply fuses the audio and visual information at the level of the input features via
\begin{equation} \label{eq:concat}
\textbf{\textrm{o}}_t=[(\textbf{\textrm{o}}_t^\mathrm{A})^T,(\textbf{\textrm{o}}_t^\mathrm{VS})^T,(\textbf{\textrm{o}}_t^\mathrm{VA})^T]^T.
\end{equation}
Superscript $T$ denotes the transpose. 
\subsection{Dynamic stream weighting}
\label{DSW}
Stream weighting is an effective method to fuse different streams. It is a solution to the problem that the various streams may be reliable and informative in very different ways. 
Hence, a number of works employ the strategy of weighting different modalities~\cite{gurban2008dynamic,heckmann2002noise,nefian2002dynamic}. Many utilize static weights; for example in~\cite{yang2005multimodal}, audio and video speech recognition models are trained separately and the DNN state posteriors of all modalities are combined by constant stream weights $\lambda^i$ according to
\begin{equation} \label{stateequalfusion}
\textnormal{log }\widetilde{p}(s |\textbf{o}_t)=\sum_{i}^{}\lambda^i\cdot \textrm{log }{p}(s |\textbf{o}_t^{i}).
\end{equation}
Here, $\textrm{log }{p}(s |\textbf{\textrm{o}}_t^{i})$ is the log-posterior of state $s$ in stream $i$ at time $t$ and $\textnormal{log}\,\widetilde{p}(s |\textbf{\textrm{o}}_t)$ is its estimated combined log-posterior.

The problem of weight determination, however, cannot be neglected~\cite{kankanhalli2006experiential}. It is clear that in good lighting conditions, the visual information may be more useful, while audio information is most beneficial in frames with a sufficiently high SNR. Therefore,  the  weight  should  be  dynamically estimated to obtain optimal fusion results. As a baseline approach, we therefore consider \emph{dynamic} stream weighting, which implements this idea. 
Specifically, we use dynamic stream weighting as described in~\cite{yu2020multimodal} as the baseline. Here, the DNN state posteriors of all modalities are combined by estimated dynamic weights according to
\begin{equation} \label{statefusion}
\textnormal{log }\widetilde{p}(s |\textbf{o}_t)=\sum_{i}^{}\lambda_t^i\cdot \textrm{log }{p}(s |\textbf{o}_t^{i}).
\end{equation}
The stream weights $\lambda_t^i$ are estimated by a feedforward network from a set of reliability measures, introduced in detail in Sec.~\ref{reliabilitymeasures}. 

Reliability information has proven beneficial for multi-modal integration in many studies~\cite{meutzner2017improving, gurban2008dynamic, hermansky2013multistream}, where it is used to inform the integration model about the degree of uncertainty in all information streams over time. In~\cite{yu2020multimodal}, the authors also consider different criteria to train the integration model. In this paper, we use two of them as our baselines, namely the mean squared error (MSE) and the cross-entropy (CE). 

This learning-based approach to weighted stream integration can effectively and significantly improve the recognition rates in lower SNR conditions. Also, in contrast to many other stream integration strategies, such as~\cite{seymour2005new,stewart2013robust,receveur2016turbo}, 
it does not suffer from a loss of performance relative to the best single modality when the modalities differ widely in their performance, but it rather gains accuracy even from the inclusion of less informative streams. This is a feature of great importance for the case at hand, as we need to design a system that will even allow for the inclusion of the visual modality under clean conditions, where audio is far more informative than video data, without loosing---or, ideally, even still gaining---performance. 

\subsection{Oracle weighting}
\label{sec:OW}
We also compute optimal, or \emph{oracle} stream weights, as described in~\cite{yu2020multimodal}. These optimal dynamic stream weights are computed in such a way as to minimize the cross-entropy with respect to the ground-truth forced alignment information. Since a known text transcription of the test set is therefore needed in this method, it is only useful to obtain a theoretical upper performance bound for standard stream-weighting approaches. To minimize the cross-entropy, 
we use convex optimization via CVX~\cite{cvx}. 

The obtained oracle stream weights $\lambda_t^i$ are then used to calculate the estimated log-posterior through Equation~\eqref{statefusion}. As oracle stream weights yield the minimum cross-entropy between the fused posteriors and the ground-truth one-hot posteriors of the reference transcription computed by forced alignment, the corresponding results can be considered as the best achievable word error rate (WER) of a stream-weighting-based hybrid recognition system. 

\subsection{End-to-end model}
In recent years, end-to-end speech recognition has quickly gained widespread popularity. The end-to-end model predicts character sequences directly from the audio signal. Comparing the end-to-end model and the hybrid speech recognition model, the end-to-end model has a lower complexity and is more easily amenable to multi-lingual ASR. But there are also some advantages to using a hybrid model. For example, the hybrid model can be learned from and adapted to comparatively little data and it can easily integrate with task-specific WFST language models~\cite{povey2011kaldi}. Importantly for this work, hybrid models allow for integration at the level of the pseudo-posteriors, which is a place for interpretable stream integration.

Hence, in this work, we use the hybrid approach to train the single modality models. To compare the performance of our proposed system to that of end-to-end AVSR, we consider the end-to-end ``Watch, Listen, Attend and Spell'' model (WLAS)~\cite{Chung17} as a baseline. In this model, the audio and video encoders are LSTM networks. The decoder is an LSTM transducer, which fuses the encoded audio and video sequences through a dual attention mechanism. 

\section{System overview}
\label{systemoverview}

In the following, we propose an architecture that centers around a decision fusion net (DFN), which learns to combine all modalities dynamically.

\begin{figure}[htbp] 
\centering
    \includegraphics[scale=0.9]{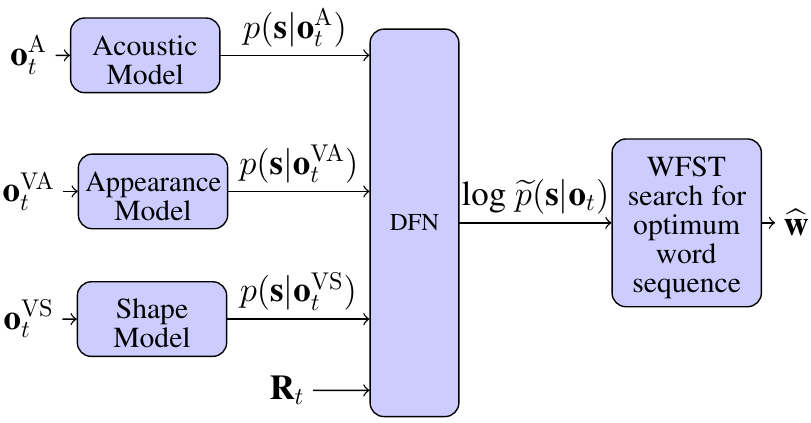}
\caption{Audio-visual fusion via the DFN, applied to one stream of audio and two streams of video features} 
\label{fig:DNN}
\end{figure}

As shown in Fig.~\ref{fig:DNN}, it bases on the state posteriors of each modality, derived from one hybrid recognition model per stream, which we consider as our representation of instantaneous feature inputs. In addition, we provide the DFN with multiple reliability indicators as auxiliary inputs, which help in estimating the multi-modal log-posteriors $\textnormal{log }\widetilde{p}(\textbf{s} |\textbf{o}_t)$ for the decoder. As mentioned above, we consider $M = 3$ single-modality models, one acoustic and two visual. The fused posterior $\textnormal{log }\widetilde{p}(\textbf{s} |\textbf{o}_t)$ is computed via
\begin{multline} \label{eq:concatprob}
\textnormal{log }\widetilde{p}(\textbf{s} |\textbf{o}_t)=\textrm{DFN}([{p}(\textbf{s} |\textbf{o}_t^{\mathrm{A}})^T,{p}(\textbf{s} |\textbf{o}_t^{\mathrm{VA}})^T, \\ {p}(\textbf{s} |\textbf{o}_t^{\mathrm{VS}})^T,\textbf{R}_t^T]^T).
\end{multline}
Here, ${p}(\textbf{s} |\textbf{o}_t^{\mathrm{A}})$, ${p}(\textbf{s} |\textbf{o}_t^{\mathrm{VA}})$ and ${p}(\textbf{s} |\textbf{o}_t^{\mathrm{VS}})$ are the state posteriors of the audio model, and of the appearance-based, and a shape-based video model, respectively. $\textbf{R}_t$ is a vector composed of the reliability measures at time $t$, which we describe in Sec.~\ref{reliabilitymeasures}. As an alternative to the posteriors of each stream, we have also considered a fusion of the log posteriors $\textnormal{log}\;{p}(\textbf{s} |\textbf{o}_t^{\mathrm{i}})$, but settled on the linear posteriors due to a better model convergence. 
\vspace{-1mm}
\vspace{-1mm}

DFN training is then performed 
with the cross-entropy loss
\begin{equation} \label{ce}
\mathcal{L}_\textnormal{CE}= -\frac{1}{T}\sum_{t=1}^{T} \sum_{s=1}^{S}p^*(s | \textbf{\textrm{o}}_t)\cdot \textnormal{log }\widetilde{p}(s |\textbf{\textrm{o}}_t).
\end{equation}
Here, $p^*(s | \textbf{\textrm{o}}_t)$ is the target state probability of state $s$, obtained by the forced alignment for the clean acoustic training data. The estimated vector of log-posteriors $\textnormal{log }\widetilde{p}(\textbf{s} |\textbf{o}_t)$ is obtained from Eq.~\eqref{eq:concatprob}. Finally, the decoder uses these estimated log-posteriors to find the optimum word sequence by a graph search through the decoding graph~\cite{mohri2008speech}.

\section{Reliability measures}
\label{reliabilitymeasures}
To support the estimation of the dynamic stream weights, we extract a range of model-based and signal-based reliability measures (see Tab.~\ref{table:RMS}), generally computed as in~\cite{yu2020multimodal}. All of these reliability indicators are used in the dynamic stream weighting baseline as well as in our proposed model.

\begin{table}[htbp]
\centering
    \caption{Overview of reliability measures}
\label{table:RMS}
    \footnotesize
    \setlength\tabcolsep{2.0pt}    
    
\renewcommand{\arraystretch}{1.2}
\begin{tabular}{|c|c|l|}\hline
\multirow{3}{*}{Model-based }& \multicolumn{2}{c|}{\multirow{2}{*}{Signal-based }}\\& \multicolumn{2}{c|}{}\\ 
\cline{2-3} 
& Audio-based & \multicolumn{1}{c|}{Video-based } \\ \hline
\multicolumn{1}{|l|}{\begin{tabular}[c]{@{}l@{}}\tabitem Entropy\\ \tabitem Dispersion \\ \tabitem Posterior difference\\ \tabitem Temporal divergence\\ \tabitem Entropy and \\ \ \ \ dispersion ratio\end{tabular}} & \multicolumn{1}{l|}{\begin{tabular}[c]{@{}l@{}}\tabitem MFCC\\  \tabitem $\Delta$MFCC\\ \tabitem SNR\\ \tabitem $f_0$ \\ \tabitem $\Delta f_0$ \\  \tabitem voicing probability \end{tabular}} & \begin{tabular}[c]{@{}l@{}}\tabitem Confidence\\ \tabitem IDCT\\ \tabitem Image \\ distortion\end{tabular} \\ \hline
\end{tabular}
\end{table}

The model-based measures are entropy, dispersion, posterior difference, temporal divergence, entropy- and dispersion-ratio. All model-based measures are computed from the log-posterior outputs of their respective single-modality models, $\textnormal{log }{p}(\textbf{s} |\textbf{o}_t)$. 

Signal-based reliability measures for the audio data comprise the first 5 MFCC coefficients with their temporal derivatives $\Delta$MFCC, again as in~\cite{yu2020multimodal}. The SNR is strongly related to the intelligibility of an audio signal. However, due to the realistic, highly non-stationary environmental noises (discussed in Sec.~\ref{setup}) used in data augmentation and testing, conventional SNR estimation algorithms are not showing a robust performance. Instead, therefore, the deep learning approach DeepXi~\cite{nicolson2019deep} is used to estimate the frame-wise SNR.

The pitch $f_0$ and its temporal derivative, $\Delta f_0$, are also used as reliability indicators. It has been shown that high pitch of a speech signal negatively affects the MFCC coefficients~\cite{dharanipragada2006robust}, 
due to insufficient smoothing of the pitch harmonics in the speech spectrum by the filterbank. 

The probability of voicing~\cite{ghahremani2014pitch} is used as an additional cue. It is computed from the Normalized Cross-Correlation Function (NCCF) values for each frame.

OpenFace~\cite{amos2016openface} is used for face detection and facial landmark extraction. This allows us to use the confidence of the face detector in each frame as an indicator of the visual feature quality. The other signal-based video reliability measures, the Inverse Discrete Cosine Transform (IDCT), and the image distortion estimates, are the same as in~\cite{yu2020multimodal}.

\section{Experimental Setup}
\label{setup}

\subsection{Dataset} \label{dataset}

The Oxford-BBC Lip Reading Sentences (LRS2) corpus is used for all experiments. It contains more than 144k sentences from British television. Table~\ref{table:dataset} gives an overview of the dataset size and partitioning. The pre-train set is usually used in AVSR tasks for video or audio-visual model pretraining. In this work, we combine the pre-train and training set to train all acoustic, visual, and AV models.

\begin{table}[htbp]
    \centering
    \caption{Size of subsets within the LRS2 Corpus}
\label{table:dataset}
    \setlength\tabcolsep{2.0pt}
    \centering
\begin{tabular}{|c|c|c|c|}
\hline
Subset                  & Utterances    & Vocabulary  &\begin{tabular}[c]{@{}l@{}}Duration\\ {[}hh:mm{]}\end{tabular} \\ \hline
pre-train set & 96,000  & 41,000 &196:25\\ 
training set & 45,839  & 17,660 &28:33\\
validation set     & 1,082   & 1,984  &00:40\\ 
test set  & 1,243   & 1,698  & 00:35 \\ \hline
\end{tabular}
\end{table}

For the AVSR task, the results are often dominated by the acoustic model. To analyze the performance in different noisy environments and to counter the audio-visual model imbalance, we add acoustic noise to the LRS2 database. The ambient subset of the MUSAN corpus~\cite{musan2015} is used as the noise source. It contains noises such as wind, footsteps, paper rustling, rain, as well as indistinct crowd noises. Seven different SNRs are selected randomly, from -9 dB to 9 dB in steps of 3 dB. 
We also generated data for a far-field AVSR scenario. As the LRS2 database does not contain highly reverberant data, we artificially reverberate the acoustic data by convolutions with measured impulse responses. These impulse responses also come from the MUSAN corpus. Both types of augmentation use Kaldi's Voxceleb example recipe.

\subsection{Feature extraction} \label{features}
The audio model uses 40 log Mel features together with two pitch features ($f_0$, $\Delta f_0$) and the probability of voicing, yielding 43-dimensional feature vectors. The audio features are extracted with 25~ms frame size and 10~ms frameshift. The video features are extracted per frame, i.e., every 40~ms. The video appearance model (VA) uses 43-dimensional IDCT coefficients of the grayscale mouth region of interest (ROI) as features. The video shape model (VS) is based on the 34-dimensional non-rigid shape parameters described in~\cite{amos2016openface}.

Since the audio and video features have different frame rates, Bresenham\textquotesingle s algorithm~\cite{sproull1982using} is used to align the video features before training the visual models. This algorithm gives the best first-order approximation for aligning audio and video frames given only a difference in frame rates.


\subsection{Implementation details} \label{impdetails}
All our experiments are based on the Kaldi toolkit~\cite{povey2011kaldi}. As mentioned in Section~\ref{dataset}, both pre-train and training sets are used together to train the acoustic and visual models. The initial HMM-GMM training follows the standard Kaldi AMI recipe, namely, monophone training followed by triphone training. A linear discriminant analysis (LDA) is applied to a stacked context of features to obtain discriminative short-term features. Finally, speaker adaptive training (SAT) is used to compensate for speaker variability. Each step produces a better forced alignment for later network training. HMM-DNN training uses the nnet2 p-norm network~\cite{zhang2014improving} recipe, which is efficiently parallelizable. 

Once HMM-DNN training has been performed, the acoustic model DNN and two visual observation models are available. They output estimated log-posteriors $\textrm{log }{p}(\textbf{s} |\textbf{\textrm{o}}_t^{i})$ for each stream, which are then used in our proposed DFN. 
Its input consists of all stream-wise state-posteriors ${p}(\textbf{s} |\textbf{\textrm{o}}_t^{i})$ and the reliability measures. 

As mentioned in Section~\ref{systemoverview}, the decoder obtains the best word sequence by graph search through a decoding graph using the estimated log-pseudo-posteriors $\textnormal{log }\widetilde{p}(\textbf{s} |\textbf{o}_t)$. To ensure that all experiments and modalities search through the same decoding graph, we share the phonetic decision tree between all single modalities. Thus, the number of states for each modality is identical, specifically 3,856.

In addition, there are 41 reliability indicators, which leads to an overall input dimension of $(3 \times 3856 + 41) =$11,609. The first three hidden layers have 8,192, 4,096, and 1,024 units, respectively, each using the ReLU activation function, layer normalization (LN), and a dropout rate of 0.15. They are followed by 3 BLSTM layers with 1,024 memory cells for each direction, using tanh as the activation function. Finally, a fully connected (FC) layer projects the data to the output dimension of 3,856. A log-softmax function is applied to obtain the estimated log-posteriors. 


Early stopping is used to avoid overfitting. We check for early stopping every 7,900 iterations, using the validation set. The training process is stopped if the validation loss does not decrease for 23,700 iterations. Finally, the performance is evaluated on the test set. We performed two experiments with the proposed DFN strategy. The first uses the BLSTM-DFN, exactly as described above, while the second is an LSTM-DFN, replacing the BLSTM layers by purely feed-forward LSTMs.

The learning rate is initialized to $5\times10^{-4}$ and reduced by 20\% whenever the validation loss does not decrease in early stopping checking. The batch size is set to 10. DNN training utilizes the PyTorch library~\cite{paszke2019pytorch} with the ADAM optimizer. 


\section{Results}
\label{results}
\begin{figure}[htbp]
    \centering
    \includegraphics[scale = 0.5]{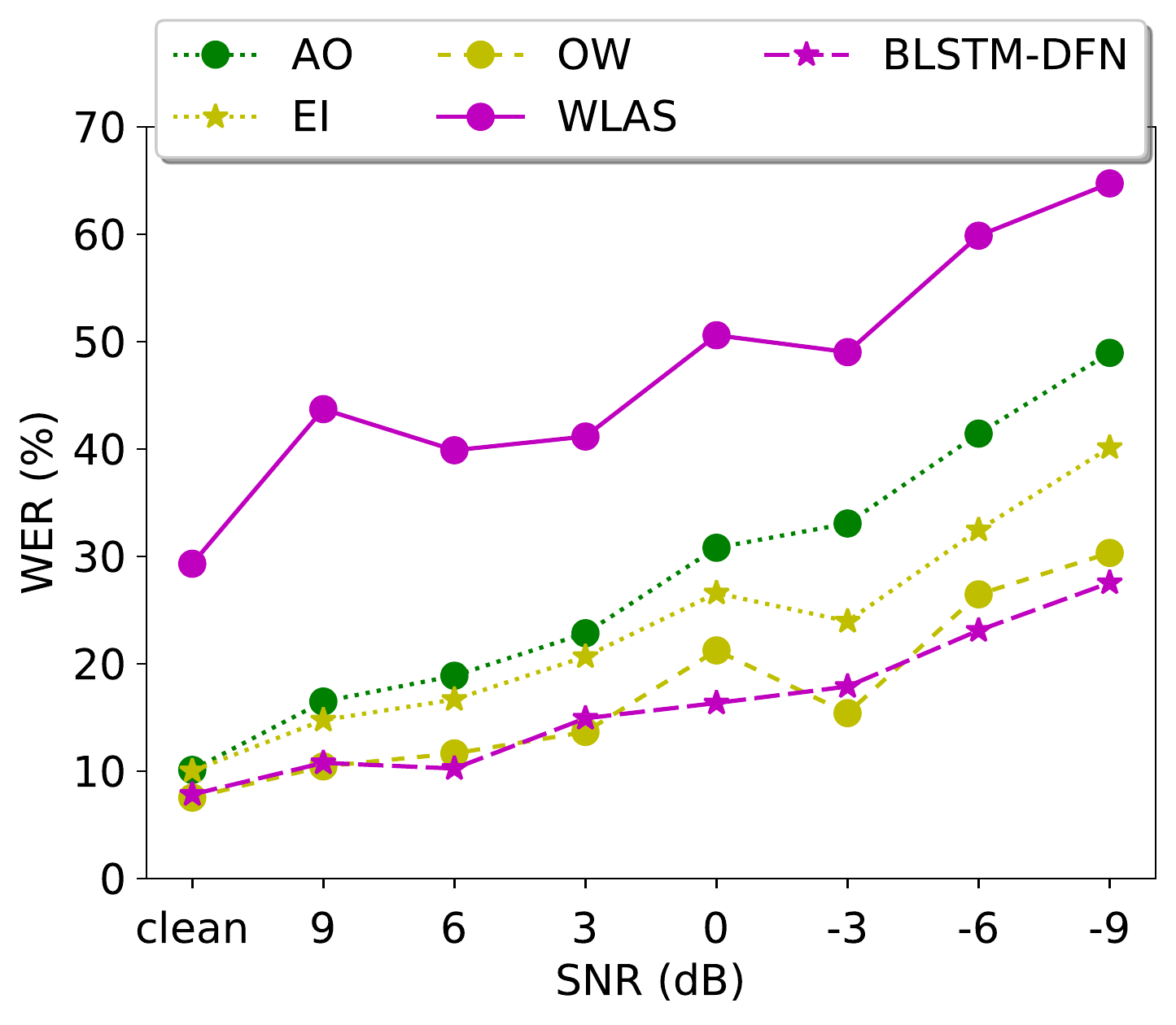}
    \caption{WER (\%) on the test set of the LRS2 corpus.}
\label{fig:baseline}
\end{figure}

In this section, we compare the performance of all baseline models and fusion strategies. Figure~\ref{fig:baseline} gives an overview of the results of the audio-only model and compares the results of all baselines and our proposed BLSTM-DFN. 



 

\vspace{-1mm}

\vspace{-1mm}

Comparing the audio-only model and the BLSTM-DFN integration, our fusion strategy is able to reduce the Word Error Rate (WER) for every SNR, even for clean acoustic data. For lower SNRs, the DFN can improve the absolute WER by over 10\%. Our proposed BLSTM-DFN is also capable of achieving better results in many cases than the--realistically unachievable---oracle weighting (OW), that is based on ground-truth transcription information of the test set and can be considered as the upper limit for the dynamic stream-weighting approach of Equation \eqref{statefusion}. The end-to-end  WLAS model is not able to improve the recognition rates comparing to the audio-only model, which may in part be due to the fact that it does not employ an explicit language model.


Table~\ref{table:results} lists all the results of our experiments under additive noise. As expected, the audio-only model (AO) has a much better performance than the video-appearance (VA) and video-shape (VS) models. The average WERs
of the visual models are over 80\%, which illustrates that lipreading is still hard in large-vocabulary tasks. We have also employed the pre-trained spatio-temporal visual front-end from~\cite{stafylakis2017combining} to extract high-level visual features, without seeing improvements. We assume that this is due to insufficient training data as well as to insufficient generalization across speakers and recording conditions.

\begin{table}[htbp]
\centering
\setlength\tabcolsep{1.75pt}
  \caption{ Word error rate (\%) on the LRS2 test set under additive noise. OW describes the realistically unachievable performance bound for dynamic, instantaneous stream weighting.}
  \begin{tabular}{@{}cccccccccc@{}}
    \toprule
    SNR  & -9    & -6    & -3    & 0     & 3     & 6     & 9     & clean & avg. \\
    \midrule
AO &48.96 &41.44 &33.07 &30.81 &22.85 &18.89 &16.49 &10.12 &27.83\\
VA &85.83 &87.00 &85.26 &88.10 &87.03 &88.44 &88.25 &88.10 &87.25\\
VS &88.11 &90.27 &87.29 &88.88 &85.88 &85.33 &88.58 &87.10 &87.68\\
\midrule
EI &40.14 &32.47 &23.96 &26.59 &20.67 &16.68 &14.76 &10.02 &23.16\\
\midrule
MSE &46.48 &37.79 &27.45 &27.47 &19.52 &16.58 &15.09 &9.42 &24.98\\
CE &45.79 &37.14 &26.32 &28.03 &19.40 &16.68 &14.76 &9.42 &24.65\\
OW &30.33 &26.47 &\textbf{15.41} &21.25 &\textbf{13.66} &11.66 &\textbf{10.45} &\textbf{7.54} &17.10\\
\midrule
WLAS &64.74& 59.87& 49.03& 50.60& 41.17& 39.89& 43.72& 29.32 &47.29\\
\midrule
LSTM-DFN  &33.30 &27.22 &21.26 &21.25 &19.17&13.97 &15.84 &10.32  &20.29\\
BLSTM-DFN &\textbf{27.55} &\textbf{23.11} &17.89 &\textbf{16.35} &14.93 &\textbf{10.25} &10.78 &7.84  &\textbf{16.09}\\
\bottomrule
\\
  \end{tabular}

\label{table:results}
\end{table}
\vspace{-2mm}

Early integration (EI) can also improve the results, 
but the improvement is not as significant as that of the proposed DFN approach. Comparing the BLSTM-DFN and the LSTM-DFN, the former shows a notable advantage in accuracy, albeit at the price of non-real-time performance. Both the LSTM-DFN and the BLSTM-DFN use recurrent layers with 1024 memory cells. As the number of parameters in a BLSTM layer are almost double that of the LSTM layer, we also trained a BLSTM-DFN using 512 memory cells per layer. The average WER of this model is 16.14\%, still better than that of the LSTM-DFN with a similar number of parameters. If we increase the number of cells for the LSTM-DFN to 2048, with the same learning rate, the model suffers from convergence issues.

The dynamic stream weighting results, using the MSE or CE loss, are better than shown in~\cite{yu2020multimodal} for three reasons. Firstly, improved reliability measures are used in this work. Secondly, \cite{yu2020multimodal} trains acoustic and visual models only on the training set, whereas here, they are trained on both the pre-train and training data. This gives a significant performance boost to the single-modality systems and also to early integration, but is not of as much added benefit to the dynamic stream-weight estimation, though the weight estimator from~\cite{yu2020multimodal} was trained on the validation set, whereas here, it is also trained on the pre-train and training sets. We assume that its relatively small performance gain is due to the limited flexibility of the composition function in dynamic stream weighting.

Comparing the average WER over all acoustic conditions, the proposed BLSTM-DFN is greatly beneficial, outperforming the not realistically achievable OW system, and surpassing all feasible stream integration approaches by a clear margin. Thus, our proposed method outperforms even optimal dynamic stream weighting and therefore provides a fundamentally superior architecture compared to instantaneous stream weighting.

\begin{threeparttable}[]
\centering
    \caption{Far-field AVSR WER (\%) on LRS2.}
\label{table:reverbresults}
    \footnotesize
\setlength\tabcolsep{4.5pt}
\begin{tabular*}{\linewidth}{ccccccccc}
\toprule
  & AO     & EI     & MSE   & CE   & OW &WLAS    & \begin{tabular}[c]{@{}c@{}}\small LSTM-\\\small DFN\end{tabular}  	& \begin{tabular}[c]{@{}c@{}}\small BLSTM-\\\small DFN\end{tabular}\\
\midrule
 &23.61 &19.15 & 19.54 & 19.44 &\textbf{12.70} & 44.24 &15.67 &15.28 \\

\bottomrule
\\
\end{tabular*}
\end{threeparttable}

We also checked the case of far-field AVSR by using data augmentation to produce artificially reverberated speech, see Table~\ref{table:reverbresults} for the results. The BLSTM-DFN still outperforms the other fusion strategies, but it is not as close to the OW. We suspect the reason is an insufficient amount of reverberant acoustic training signals. 

Overall, it can be concluded that the introduced DFN is generally superior to instantaneous dynamic stream weighting. The latter can be considered as fusion at the \emph{frame level}. Frame-by-frame, it sums log-posteriors of each stream in a weighted fashion. Hence, its estimated combined log-posterior is a linear transformation of the single-modality log-posteriors. In contrast, the DFN can be considered as a cross-temporal fusion strategy at the \emph{state level}, as the combined log-posterior is estimated through a non-linear transformation with memory. This allows for a more accurate estimation, in which the BLSTM-DFN gives an added advantage to the LSTM-DFN, since it has access to both past and future contextual information. In this work, the BLSTM-DFN shows a relative WER reduction of 42.18\% compared to the audio-only system, while the LSTM-DFN yields a relative WER improvement of 27.09\%, showing the benefit of being able to lipread even for noisy LVCSR.

\section{Conclusion}
\label{conclusion}
There are still many difficulties for large-vocabulary speech recognition under adverse conditions, but the fusion of acoustic and visual information can bring a significant benefit to these challenging and interesting tasks. In this paper, we propose a new architecture, the decision fusion net (DFN), in which we consider state posteriors of acoustic and visual models as appropriate stream representations for fusion. These are combined by the DFN, which uses stream reliability indicators to estimate the optimal state-posteriors for hybrid speech recognition. It comes in two flavors, a BLSTM-DFN with optimal performance, as well as an LSTM-DFN, which provides the option of real-time decoding.

We compare the performance of our proposed model to early integration as well as to conventional dynamic stream weighting models. In experimental results on noisy as well as on reverberant data, our proposed model shows significant improvements, with the BLSTM version giving a relative word-error-rate reduction of 42.18\% over audio-only recognition, and outperforming all baseline models. The hybrid architecture with the proposed DFN clearly outperforms the end-to-end WLAS model, which we attribute to its factorization of stream evaluation, stream integration, and subsequent, language-model-supported, search.  It is worth mentioning that, on average, the hybrid DFN model is even superior to a hybrid model with \emph{oracle} stream weighting, which is an interesting result on its own, given that the latter provides a theoretical upper bound for instantaneous stream weighting approaches. 

The natural next goal of our work is to focus on end-to-end audio-visual speech recognition models. Here, we are specifically interested in investigating reliability-supported fusion within the attention mechanism in CTC and transformer systems and in the possibilities that come with probabilistic intermediate representations for these architectures.

\bibliographystyle{IEEEtran}
\bibliography{IEEEabrv,ref}

\end{document}